\documentclass[aps,twocolumn,a4paper,floatfix,superscriptaddress,prl]{revtex4-1}
\usepackage{graphicx}
\usepackage[toc,page]{appendix}
\usepackage{braket}
\usepackage{epsfig,graphicx,times}
\usepackage{amstext}
\usepackage{amsmath}
\usepackage{amssymb}
\usepackage{mathrsfs}
\usepackage{dcolumn}
\usepackage{bm}
\usepackage[tight]{subfigure}
\usepackage[colorlinks,linkcolor=blue,anchorcolor=blue,citecolor=blue,urlcolor=blue]{hyperref}
\usepackage{color}
\usepackage{siunitx}
\usepackage{appendix}
\usepackage{placeins}
\usepackage{textcomp}
\usepackage{bbold}
\usepackage{float}
\usepackage{verbatim}
\usepackage{minitoc}
\usepackage[dvipsnames]{xcolor}

\newcommand{\prlsection}[1]{\emph{#1.}---}
\usepackage{soul}
\usepackage[framemethod=tikz]{mdframed}
\usepackage{lipsum}
\usepackage{epstopdf}

\begin{document}

\preprint{APS/123-QED}

\title{Strong Molecule-Light Entanglement with Molecular Cavity Optomechanics}

\author{Hong-Yun Yu}
\affiliation{College of Advanced Interdisciplinary Studies, NUDT, Changsha 410073, P.R.China}
\author{Ya-Feng Jiao}
\email{yfjiao@zzuli.edu.cn}
\affiliation{School of Electronics and Information, Zhengzhou University of Light Industry, Zhengzhou 450001, P.R.China}
\affiliation{Academy for Quantum Science and Technology, Zhengzhou University of Light Industry, Zhengzhou 450001, P.R.China}

\author{Jie Wang}
\affiliation{Key Laboratory of Low-Dimensional Quantum Structures and Quantum Control of Ministry of Education, Department of Physics and Synergetic Innovation Center for Quantum Effects and Applications, Hunan Normal University, Changsha 410081, P.R.China}
\author{Feng Li}
\affiliation{Key Laboratory for Physical Electronics and Devices of the Ministry of Education \& Shaanxi Key Laboratory of Information Photonic Technique, School of Electronic Science and Engineering, Faculty of Electronic and Information Engineering, Xi’an Jiaotong University, Xi’an 710049, P.R.China}
\author{Bin Yin}
\affiliation{Key Laboratory of Low-Dimensional Quantum Structures and Quantum Control of Ministry of Education, Department of Physics and Synergetic Innovation Center for Quantum Effects and Applications, Hunan Normal University, Changsha 410081, P.R.China}

\author{Tian Jiang}
\affiliation{Institute for Quantum Science and Technology, College of Science, NUDT, Changsha 410073, P.R.China}
\affiliation{College of Advanced Interdisciplinary Studies, NUDT, Changsha 410073, P.R.China}
\affiliation{Hunan Research Center of the Basic Discipline for Physical States,  Changsha 410073, P.R.China}

\author{Qi-Rui Liu}
\affiliation{College of Advanced Interdisciplinary Studies, NUDT, Changsha 410073, P.R.China}

\author{Hui Jing}
\email{jinghui73@foxmail.com}
\affiliation{Key Laboratory of Low-Dimensional Quantum Structures and Quantum Control of Ministry of Education, Department of Physics and Synergetic Innovation Center for Quantum Effects and Applications, Hunan Normal University, Changsha 410081, P.R.China}
\affiliation{Institute for Quantum Science and Technology, College of Science, NUDT, Changsha 410073, P.R.China}

\author{Ke Wei}
\email{weikeaep@163.com}
\affiliation{Institute for Quantum Science and Technology, College of Science, NUDT, Changsha 410073, P.R.China}
\date{\today}

\begin{abstract}
We propose a molecular optomechanical platform to generate robust entanglement among bosonic modes—photons, phonons, and plasmons—under ambient conditions. The system integrates an ultrahigh-Q whispering-gallery-mode (WGM) optical resonator with a plasmonic nanocavity formed by a metallic nanoparticle and a single molecule. This hybrid architecture offers two critical advantages over standalone plasmonic systems: 
(i) Efficient redirection of Stokes photons from the lossy plasmonic mode into the long-lived WGM resonator, and 
(ii) Suppression of molecular absorption and approaching vibrational ground states via plasmon-WGM interactions. 
These features enable entanglement to transfer from the fragile plasmon-photon subsystem to a photon-phonon bipartition in the blue-detuned regime, yielding robust stationary entanglement resilient to environmental noise. Remarkably, the achieved entanglement surpasses the theoretical bound for conventional two-mode squeezing in certain parameter regimes. Our scheme establishes a universal approach to safeguard entanglement in open quantum systems and opens avenues for noise-resilient quantum information technologies.
\end{abstract}

\maketitle

\prlsection{\label{sec:1-1} Introduction}
Cavity optomechanical (COM) systems, which exploit the radiation-pressure-induced nonlinear interactions between electromagnetic fields and vibrational modes \cite{aspelmeyerCavityOptomechanics2014}, have emerged as a cornerstone for exploring quantum phenomena in macroscopic systems.
These platforms have enabled groundbreaking advances in ultra-sensitive sensing \cite{zhaoWeakforceSensingSqueezed2020}, laser cooling \cite{wilson-raeCavityassistedBackactionCooling2008}, and quantum information processing \cite{millenOptomechanicsLevitatedParticles2020}. 
Recently, the integration of plasmonic nanostructures with molecular vibrations has expanded the scope of COM physics into the realm of molecular optomechanics \cite{roelliMolecularCavityOptomechanics2016,roelliNanocavitiesMolecularOptomechanics2024,schmidtQuantumMechanicalDescription2016,schmidtLinkingClassicalMolecular2017,estebanMolecularOptomechanicsApproach2022}.
By leveraging surface-enhanced Raman scattering (SERS) in metallic nanoparticles, this framework provides a quantum-mechanical description of light-matter interactions at the single-molecule level, revealing phenomena such as phonon parametric amplification \cite{jakobGiantOptomechanicalSpring2023,xuPhononicCavityOptomechanics2022,lombardiPulsedMolecularOptomechanics2018}, optomechanical frequency conversion \cite{chenContinuouswaveFrequencyUpconversion2021,xomalisDetectingMidinfraredLight2021}, and vibrationally mediated photon blockade \cite{abutalebiSinglephotonGenerationRoom2024,moradikalardePhotonAntibunchingSinglemolecule2025}.
Such effects are not only pivotal for quantum photochemistry \cite{liuSurfacePlasmonicCatalysis2022} and nanoscale heat transfer \cite{ashrafiOptomechanicalHeatTransfer2019} but also establish molecular COM systems as versatile platforms for encoding quantum information via spin, dipole, or vibrational degrees of freedom \cite{carrettaPerspectiveScalingQuantum2021,ruttleyLonglivedEntanglementMolecules2025,cornishQuantumComputationQuantum2024}.
Critically, molecular COM systems offer three transformative advantages over conventional optomechanical architectures: (i) subwavelength plasmonic field confinement amplifies single-photon coupling strengths by orders of magnitude \cite{estebanMolecularOptomechanicsApproach2022,liBoostingLightMatterInteractions2024a}, (ii) high-frequency molecular vibrations suppress thermal decoherence, promising ambient temperature quantum phenomena with solid-state COM devices \cite{chenContinuouswaveFrequencyUpconversion2021,jakobGiantOptomechanicalSpring2023,abutalebiSinglephotonGenerationRoom2024}, and (iii) intrinsic molecular degrees of freedom enable novel pathways for quantum information processing, such as spin-based qubit encoding \cite{carrettaPerspectiveScalingQuantum2021,chiesaMolecularNanomagnetsViable2024} or dipole mediated quantum computation \cite{ruttleyLonglivedEntanglementMolecules2025,cornishQuantumComputationQuantum2024}.

Quantum entanglement, a hallmark of nonclassical correlations, serves as the backbone of quantum technologies ranging from secure communication to distributed quantum networks \cite{weedbrookGaussianQuantumInformation2012,bennettMixedstateEntanglementQuantum1996,raussendorfOneWayQuantumComputer2001,laddQuantumComputers2010,xiaDemonstrationReconfigurableEntangled2020,degenQuantumSensing2017}.
While entanglement has been demonstrated across diverse platforms, including photons \cite{zhaoExperimentalDemonstrationFivephoton2004,dadaExperimentalHighdimensionalTwophoton2011}, trapped ions \cite{mazzantiTrappedIonQuantum2021,haffnerQuantumComputingTrapped2008}, superconducting circuits \cite{clarkeSuperconductingQuantumBits2008}, and macroscopic mechanical oscillators \cite{kotlerDirectObservationDeterministic2021a,mercierdelepinayQuantumMechanicsFree2021}, its fragility to environmental decoherence remains a persistent challenge.
In COM systems, entanglement between photons and phonons is typically generated via nonlinear two-mode squeezing interactions \cite{vitaliOptomechanicalEntanglementMovable2007,palomakiEntanglingMechanicalMotion2013}, but its stationary value is fundamentally bounded by intracavity loss, thermal noise and system instability \cite{wangReservoirEngineeredEntanglementOptomechanical2013,genesRobustEntanglementMicromechanical2008}.
Recent efforts to enhance robustness have employed dynamical modulation \cite{wangMacroscopicQuantumEntanglement2016,huManifestationClassicalNonlinear2019,huangQuadraturesqueezedLightOptomechanical2018}, reservoir engineering \cite{zhangEnhancedOptomechanicalEntanglement2020,wangReservoirEngineeredEntanglementOptomechanical2013,liaoReservoirengineeredEntanglementHybrid2018}, 
quantum interference \cite{tianRobustPhotonEntanglement2013}
and synthetic gauge fields \cite{liuPhasecontrolledAsymmetricOptomechanical2023}, yet these strategies often require stringent experimental conditions or fail to address the limitations of lossy plasmonic modes in single-molecule COM systems. 
Although ensemble-based molecular COM platforms have achieved room-temperature entanglement through collective coupling \cite{huangCollectiveQuantumEntanglement2024a,emaleNonreciprocalEntanglementMolecular2025}, the deterministic generation and preservation of entanglement in a single-molecule architecture—a critical step toward scalable quantum technologies—remains elusive.
This gap stems from the inherent trade-off between plasmonic field enhancement and radiative losses, as well as the difficulty in suppressing effective thermal phonon populations.

\begin{figure*}
\centering
    \includegraphics{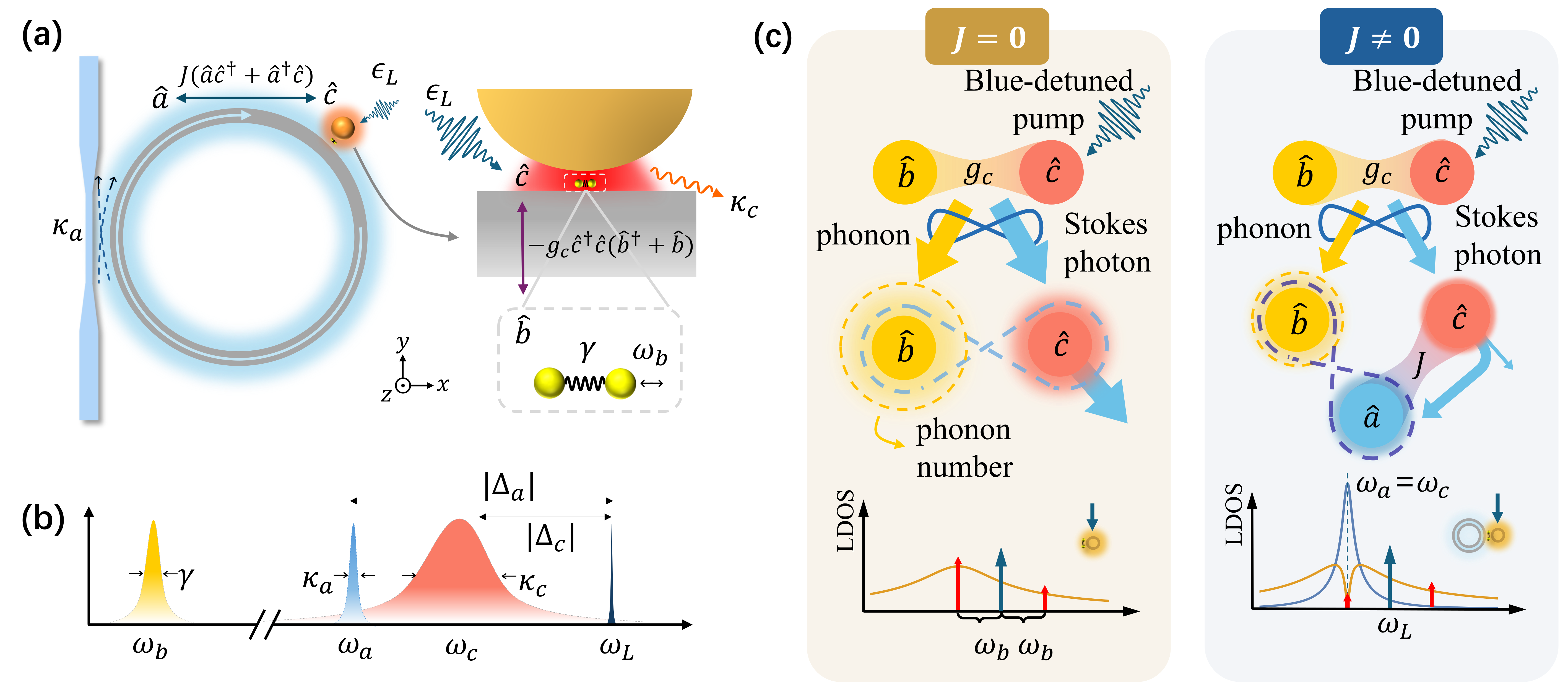}
    \caption{\label{fig:Fig1}(a) Schematic diagram of single-molecule optomechanical system with WGM (denoted as $\hat a$ and supported by microdisk cavity). The molecular optomechanical system consists of a plasmonic mode (denoted as $\hat c$, supported by metallic nanoparticle) and a molecular vibration mode (denoted as $\hat b$). The system is pumped from plasmonic mode with amplitude $\varepsilon_L$. $J$ is the coupling strength between WGM and plasmonic mode; $g_c$ is the single-photon coupling strength between plasmon and molecular vibration mode. (b) Frequency spectrum of the system. The molecular vibration mode has frequency $\omega_b$ and decay $\gamma$. The WGM has frequency $\omega_a$ and decay $\kappa_a$. The plasmonic mode has frequency $\omega_c$ and decay $\kappa_c$. (c) Schematic diagram of bipartite entanglement generation and transformation. The bottom line plots in the two boxes are schematic diagram of local mode density of states (LDOS) of plasmonic mode (yellow) and WGM (blue) and Stokes/anti-Stokes amplitude (red arrows).}
\end{figure*}

Here, we address these challenges by coupling a plasmonic nanocavity (hosting a single molecule and metallic nanoparticle) to a high-Q whispering-gallery-mode (WGM) resonator. The WGM resonator redirects Stokes photons from the lossy plasmonic mode into a low-loss optical channel, suppressing molecular absorption and approaching vibrational ground states.
Under blue-detuned driving, entanglement generated in the plasmon-photon subsystem is transferred to a photon-phonon bipartition via linear coupling.
Remarkably, this mechanism yields stationary entanglement that surpasses theoretical limits for two-mode stationary intracavity entanglement and exhibits resilience to both thermal fluctuations and plasmonic dissipation. 

\prlsection{\label{sec:sys} The System}
In Figure\,\ref{fig:Fig1}(a), We propose a hybrid molecular cavity optomechanical system combining a plasmonic nanocavity with a microdisk whispering-gallery-mode (WGM) resonator. The nanoparticle-on-mirror (NPoM) plasmonic cavity, containing a biphenyl-4-thiol molecule, supports both a plasmonic mode ($\omega_c,\kappa_c$) and a molecular vibration mode ($\omega_b,\gamma$). The WGM resonator sustains an optical mode ($\omega_a, \kappa_a$) evanescently coupled to the plasmonic mode with strength $J$.
Plasmon-phonon interaction emerges via radiation pressure with single-photon coupling rate $g_c$.
This architecture leverages the plasmonic field’s subwavelength confinement over a broad spectrum to enhance single-photon optomechanical coupling ($g_c$) 
while utilizing the WGM’s low dissipation ($\kappa_a$) to preserve quantum coherence.

Under coherent driving of the plasmonic mode with laser power $P$ and frequency $\omega_L$, the system’s dynamics are governed by the Hamiltonian in the rotating frame:
\begin{align}
\hat H=~&\hbar \Delta_a \hat{a}^\dagger \hat{a}+\hbar \Delta_c \hat{c}^\dagger \hat{c}+\hbar \omega_b \hat{b}^\dagger \hat{b}
-\hbar g_c \hat{c}^\dagger \hat{c}(\hat{b}^\dagger+\hat{b})\nonumber\\
&+\hbar J(\hat{a}  \hat{c}^\dagger+\hat{a}^\dagger  \hat{c})
+i\hbar \varepsilon_L(\hat{c}^\dagger-\hat{c}),
\label{eq:Hami}
\end{align}  
where $\Delta_{a,c}=\omega_{a,c}-\omega_L$ are detunings, $\hat{a},\hat b,\hat c$ ($\hat{a}^\dagger$,$\hat{b}^\dagger$,$\hat{c}^\dagger$) are the annihilation (creation) operators of the optical mode, molecular vibration mode and plasmonic mode, respectively,
and $\varepsilon_L=\sqrt{2\kappa_c P/\hbar \omega_L}$ is the laser amplitude. The nonlinear term $-\hbar g_c \hat{c}^\dagger \hat{c}(\hat{b}^\dagger+\hat{b})$ describes radiation-pressure-mediated plasmon-phonon coupling, while the beam-splitter term $\hbar J(\hat{a}  \hat{c}^\dagger+\hat{a}^\dagger  \hat{c})$ enables photon exchange between the WGM and plasmonic modes.
Here, we neglect the WGM-molecular vibration coupling and the increase of plasmonic mode volume due to plasmon-WGM coupling \cite{doelemanAntennaCavityHybrids2016}, because the high-Q WGM resonator is specifically designed to preserve Stokes photons at $\Delta_a=-\omega_b\sim\kappa_c$ (Fig.\,\ref{fig:Fig1}(c)). At this detuning, the pump laser ($\omega_L$) is effectively decoupled from WGM, making the radiation pressure from the WGM field and electric field distribution in WGM resonator negligible.

Steady state bipartite entanglement can be calculated from the $6\times6$ covariance matrix (CM) $V$ of fluctuation-quadrature operators of our three bosonic modes, which satisfies the Lyapunov equation \cite{vitaliOptomechanicalEntanglementMovable2007}: $AV+VA^T=-D$. 
Here, $A$ is the drift matrix, encoding the mode frequencies $\Delta_{a},\Delta_c,\omega_b$, dissipation rates $\kappa_{a},\kappa_c,\gamma$, plasmon-photon coupling strength $J$ and the effective plasmon-phonon coupling strength $G=2g_cc_s$ (see Eq.\,(S8)), $c_s$ is the mean amplitude of plasmonic mode $\hat c$ under laser driving.
$D=\mathrm{diag}[\kappa_a,\kappa_a,\gamma(2 \bar n+1),\gamma(2 \bar n+1),\kappa_c,\kappa_c]$ is the diffusion matrix and $\bar n$ is the thermal phonon number. 
Benefits from the high frequency nature of the molecular vibration ($\sim \mathrm{THz}$), $\bar n$ is severely inhibited to $\le0.01$ even at room temperature.
Bipartite entanglement between any two modes ($n,m$ with $m(n)=\hat a,\hat b,\hat c$) is quantified with logarithmic negativity $E_{Nm|n}=\mathrm{max}\lbrack 0,-\ln(2v^-_{m,n})\rbrack$ \cite{adessoExtremalEntanglementMixedness2004}, where $v^-_{m,n}$ is  the minimum symplectic eigenvalue of the partial transpose of a reduced 4 × 4 CM $V_{m,n}$, slicing from $V$ with selected rows and columns of mode $m$ and $n$.
The definition of $E_{Nm|n}$ indicates that the selected bipartition $m$ and $n$ gets entangled if and only if $v^{-}_{m,n}<1/2$ (i.e., $E_{Nm|n}>0$).

The system operates with experimentally feasible parameters \cite{sunRevealingPhotothermalBehavior2022,shlesingerHybridCavityantennaArchitecture2023,lombardiPulsedMolecularOptomechanics2018,chenContinuouswaveFrequencyUpconversion2021,shlesingerIntegratedMolecularOptomechanics2021,chenContinuouswaveFrequencyUpconversion2021}: 
$\omega_a/2\pi=\omega_c/2\pi=330 \,\mathrm{THz},\kappa_a/2\pi=10^{-4} \,\mathrm{THz},\kappa_c/2\pi=15 \,\mathrm{THz},\omega_b/2\pi=30 \,\mathrm{THz},\gamma/2\pi=0.01 \,\mathrm{THz},J/2\pi\in[0,1.5] \,\mathrm{THz},G/2\pi\in[0,2] \,\mathrm{THz}$ and $\bar n\approx 0.01$ for molecular vibration mode ($\sim\mathrm{THz}$) at ambient condition.
With these parameters, CM $V$ can be numerically solved via Lyapunov equation and bipartite entanglement $E_{Na|b}$ (photon-phonon) and $E_{Nc|b}$ (plasmon-phonon) can be calculated from CM $V$. Detailed deriving and calculating process refer to Sec.I of SI. 

\begin{figure}
    \centering
    \includegraphics{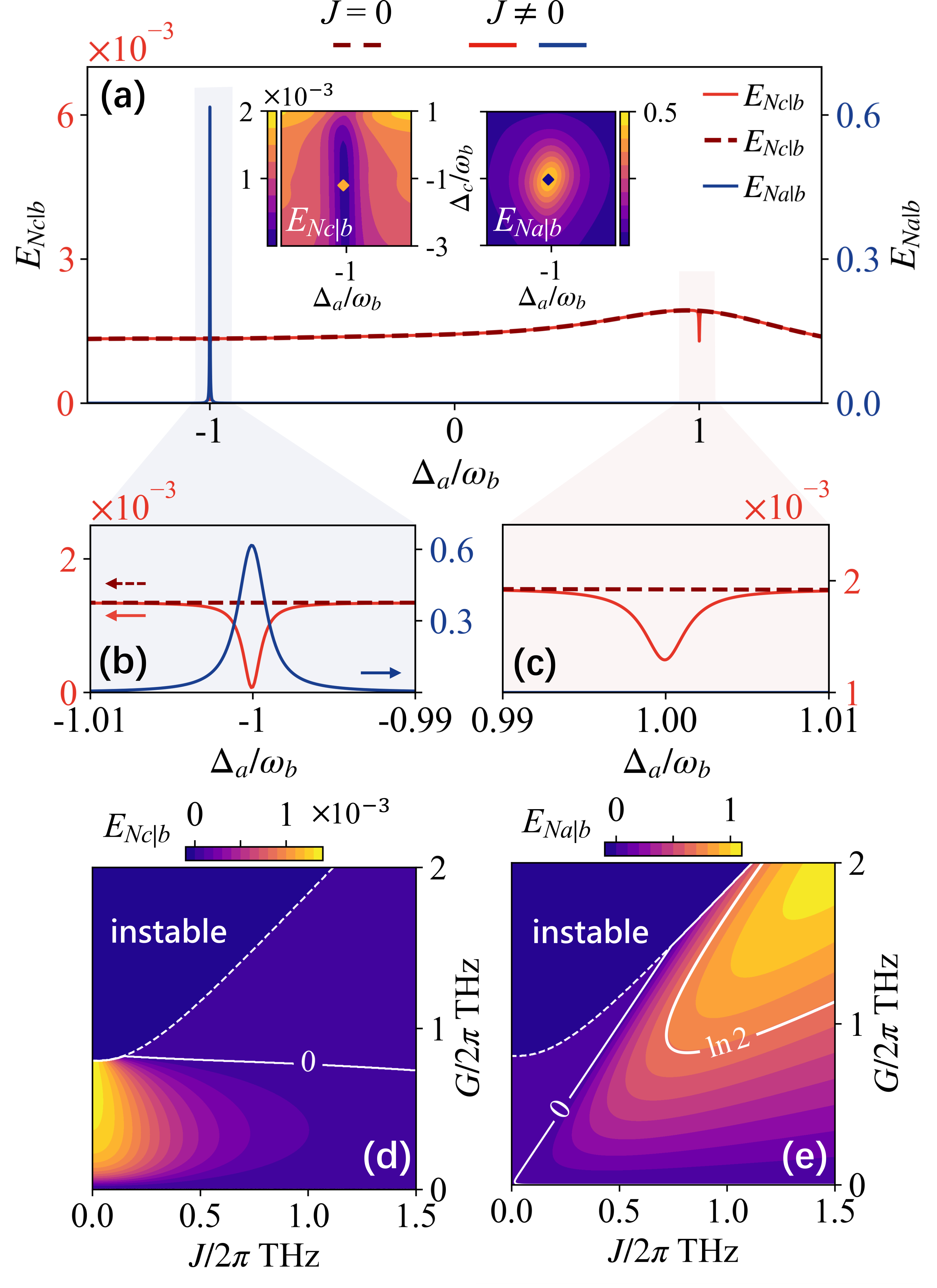}
    \caption{(a) The line plots of the logarithmic negativity $E_{Na|b}$ and $E_{Nc|b}$ versus the scaled optical detuning $\Delta_a/\omega_b$ for plasmon-photon coupling strength $J/2\pi=0.7\,\mathrm{THz}$ (solid line) and $J/2\pi=0$ (dashed line), here we take the effective plasmon-phonon coupling strength $G/2\pi=0.7\,\mathrm{THz}$ and $\Delta_a=\Delta_c$. $E_{Na|b}$ and $E_{Nc|b}$ denote the photon-phonon entanglement and the plasmon-phonon entanglement, respectively. The insets in (a) are contour-filled plots of $E_{Na|b}$ (right) and $E_{Nc|b}$ (left) as functions of scaled optical detuning $\Delta_a/\omega_b\in[-1.002,-0.998]$ and $\Delta_c/\omega_b\in[-3,1]$, with $J/2\pi=0.5\,\mathrm{THz}$, $G/2\pi=0.7\,\mathrm{THz}$.
    (b),(c) are the local magnified figures of $E_{Na|b}$ and $E_{Nc|b}$ in (a) around the COM resonances $\Delta_c/\omega_b\simeq$ (b) $-1$ and (c) $1$. 
    (d),(e) are contour-filled plots of $E_{Nc|b}$ (d) and $E_{Na|b}$ (e) as functions of $J$ and $G$ with $\Delta_a/\omega_b=\Delta_c/\omega_b=-1$. The other parameters are provided in the text.}
    \label{fig:Fig2}
\end{figure}

\prlsection{\label{sec:EnTaoff}Entanglement Trade-off}
We first demonstrate how to achieve robust entanglement under ambient conditions by transferring entanglement from the lossy plasmon-photon bipartition to the photon-phonon bipartition.
Figure\,\ref{fig:Fig2}(a) first shows a simple case when $\omega_a=\omega_c$ (i.e., $\Delta_a=\Delta_c$) and plots the logarithmic negativity $E_{Na|b}$ (photon-phonon) and $E_{Nc|b}$ (plasmon-phonon) versus the scaled optical detuning $\Delta_a/\omega_b$.
In the absence of a WGM resonator (i.e., $J=0$), 
$E_{Nc|b}$ spans a broad detuning range but remains weak (the order of $O(10^{-3})$) due to the large linewidth $\kappa_c$ of the plasmonic mode and relatively weak effective coupling strength $G$ of a single molecule.
This weak entanglement originates from down-conversion (or two-mode squeezing) interaction mediated by the nonlinear plasmon-phonon coupling $-\hbar g_c \hat{c}^\dagger \hat{c}(\hat{b}^\dagger+\hat{b})$ in Hamiltonian (\ref{eq:Hami}), 
analogous to conventional optomechanical entanglement generation. However, rapid plasmonic decoherence $\kappa_c\gg\gamma$ degrades correlations before they can stabilize, leaving $E_{Nc|b}$ extremely weak. 

Introducing finite $J$ activates a beam-splitter coupling $\hbar J(\hat{a}  \hat{c}^\dagger+\hat{a}^\dagger  \hat{c})$, which redistributes entanglement from the lossy plasmon-phonon subsystem to the high-Q photon-phonon pair. As shown in Fig.\,\ref{fig:Fig2}(b), around the Stokes sideband ($\Delta_a=\Delta_c=-\omega_b$, the blue-detuned regime), $E_{Nc|b}$ collapses sharply, and entanglement is effectively transferred to photon-phonon pair, leading to a nonzero $E_{Na|b}$ peaking at $0.616$, a value more than two orders of magnitude larger than $E_{Nc|b}$. 
Moreover, $E_{Na|b}$ and $E_{Nc|b}$ are sensitive to the variance of the optical detuning $\Delta_c$ and $\Delta_a$, and the entanglement transfer happens only with a finite interval around the resonance $\Delta_c/\omega_b=\Delta_a/\omega_b=-1$ (insets of Fig.\,\ref{fig:Fig2}(a)), which indicates that Stokes photon---generated in down-conversion process---plays a key role in entanglement transferring.
Such large entanglement in photon-phonon pair is attributed to the high-Q factor of the optical mode (long photon life) and the low thermal occupancy of the high-frequency vibration mode (low thermal noise).

\begin{figure}
    \centering
    \includegraphics{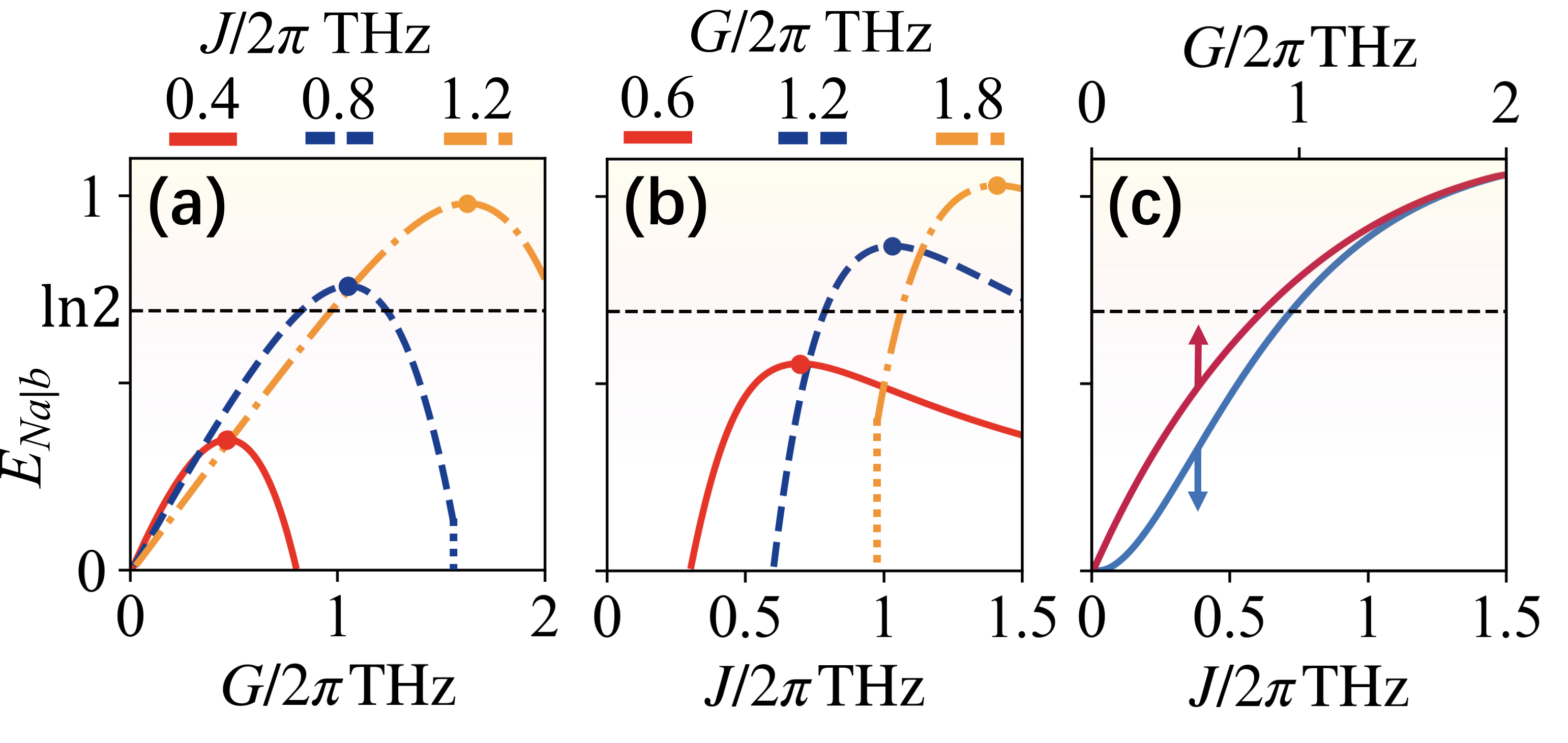}
    \caption{(a) $E_{Na|b}$ versus $G$ for different values of plasmon-photon coupling strength $J$. 
    (b) $E_{Na|b}$ versus $J$ for different values of effective plasmon-phonon coupling strength $G$. 
    (c) Local maximum of $E_{Na|b}$ versus $J$($G$) by tuning $G$($J$). All points in Fig.\,\ref{fig:Fig3} are in stable parameter region.}
    \label{fig:Fig3}
\end{figure}

The tunability of entanglement with respect to $J$ and $G$ is systematically  quantified in Figs.\,\ref{fig:Fig2}(d,e). Increasing $J$ enhances $E_{Na|b}$ while simultaneously diminishes $E_{Nc|b}$, confirming the complementary nature of the two bipartitions. 
Notably, the instability threshold $G$ rises with increasing $J$, indicating that plasmon-photon coupling can stabilize the system, allowing it sustaining stronger pump power (larger $G$) and thus larger entanglement. 
The plasmon-phonon bipartition becomes fully disentangled ($E_{Nc|b} \to 0$) as $G>2\pi\times 0.8\,\mathrm{THz}$, allowing completely isolation of the entanglement from dissipative plasmon mode.
In this way, entanglement can be separated from the production base (plasmonic hotspot) and transferred to the storage site (WGM resonator) for large, robust and long-term storage.
However, excessively large values of $J$ or $G$ degrade $E_{Na|b}$, as shown in Fig.\,\ref{fig:Fig3}(a,b).
This nonmonotonic behavior arises from two competing mechanisms, which will be further discussed in next section.
Here, over-strong $J$ suppresses two-mode squeezing interaction and the storage of Stokes photon in WGM resonator, while over-strong $G$ enhances thermal phonon noise, both of which have adverse impacts on $E_{Na|b}$. Remarkably, parameter optimization on $J$ and $G$ enables $E_{Na|b}$ to break the theoretical bound for two-mode stationary intracavity entanglement ($E_{N}^\mathrm{max}=\ln2\approx0.69$) easily, reaching $1.05$ at $J/2\pi=1.5\,\mathrm{THz},G/2\pi=2\,\mathrm{THz}$ (Fig.\,\ref{fig:Fig3}(c)).
This result is reminiscent of the recent proposals for preparing strong steady-state entanglement in three-mode optomechanical systems, which reveal that the entanglement bound can be exceeded by controlling the thermal noise of mechanical mode through reservoir engineering \cite{zhangEnhancedOptomechanicalEntanglement2020,wangReservoirEngineeredEntanglementOptomechanical2013,liaoReservoirengineeredEntanglementHybrid2018} or quantum interference \cite{tianRobustPhotonEntanglement2013}. 
The physical mechanism of our proposed scheme is fundamentally distinct from these proposals, as will be discussed in detail in the following section.

\prlsection{\label{sec:noise} Breaking the entanglement bound via photon redirection and noise suppression}
The entanglement transfer mechanism above enables a critical advance: surpassing the long-standing theoretical limit $E_{N}^\mathrm{max}=\ln2$ for two-mode stationary intracavity entanglement.
This bound arises from the two competing factors in two-mode squeezing interaction.
Specifically, correlated Stokes photon-phonon pair generated in this process (two-mode squeezing) can enhance entanglement, but overmany phonons leads to large thermal noise which suppresses entanglement.
In conventional systems, these competing effects restrict logarithmic negativity $E_N$ to sub-logarithmic values.
In our hybrid architecture, however, introducing the WGM-photon coupling $J$ activates a photon-redirection channel from the lossy plasmonic mode to the high-quality WGM resonator, which may surpass this theoretical bound.

To quantify this redirection channel, we carefully analyze the relationship between quantum entanglement and plasmonic and optical modes. Since quantum correlation is generated between phonon and Stokes photon via down-conversion (or two-mode squeezing) interaction, the amount of photon-phonon entanglement is partially determined by the extent to which Stokes photons are transferred to the optical mode. Thus, the photon-redirection can be quantified by the asymmetric response ratio of the optical ($f_a[-\omega_b]$) and plasmonic ($f_c[-\omega_b]$) modes (see Sec.II of SI for detail):
\begin{equation}
    R_{ac}\equiv\frac{f_a[-\omega_b]}{f_c[-\omega_b]}=-\frac{iJ}{\kappa_a+i(\omega_b+\Delta_a)}.
    \label{eq:ratio}
\end{equation}
In the case of low optical decay rates $\kappa_a$ (high-Q WGM), Eq.\,(\ref{eq:ratio}) predicts that the response ratio $R_{ac}\gg1$ when $J\gg\kappa_a$ and $\Delta_a/\omega_b=-1$, which is easily achieved because of the high-Q WGM resonator. 
This result indicates that at Stokes sideband, the large photon-plasmon coupling strength can redirect most of Stokes photons generated in the plasmonic mode to the WGM \cite{shlesingerIntegratedSidebandResolvedSERS2023}, 
where quantum entanglement is significantly enhanced due to the increased photon life time.

While photon-redirection enables molecular vibration mode to entangle with a high-Q optical mode, enhancing $E_{Na|b}$ by two orders of magnitude compared to $E_{Nc|b}$, this mechanism alone cannot ensure $E_{Na|b}$ exceeding the theoretical limit of $\ln2$.
Below, we demonstrate that the high-Q WGM resonator further leads to the suppression of thermal noise, which manifests itself as the ground state cooling of molecular vibration mode. When synergistically combined with photon-redirection, this dual approach allows $E_{Na|b}$ to surpass the $\ln2$ threshold, achieving a regime previously deemed technically challenging. 
We first calculate the effective stationary excitation numbers of phonon $\langle N_{\hat b}\rangle$ based on the optomechanical perturbation theory (See Sec.III of SI for detail), where $\langle N_{\hat b}\rangle<1$ signifies the ground state cooling of the vibrational mode. At Stokes sideband resonance ($\Delta_a=\Delta_c=-\omega_b$), we have~\cite{aspelmeyerCavityOptomechanics2014,liuReviewCavityOptomechanical2013}:
\begin{equation}
    \langle N_{\hat b} \rangle = \frac{2\bar n \gamma +A_+}{2\gamma-A_++A_-},
    \label{eq:number}
\end{equation}
where $A_+=G^2/2\times\kappa_a/(J^2+\kappa_a\kappa_c)$ is the light absorption rate, also the rate of generating correlated Stokes photon-phonon pair,
and $A_-\approx G^2/2\times\kappa_c/(4\omega_b^2+\kappa_c^2)$ is the light emission rate, which is insensitive to WGM resonator's coupling.
For a typical value of $\kappa_a/2\pi=10^{-4}\,\mathrm{THz}$ and $\kappa_c/2\pi=15\,\mathrm{THz}$, the light absorption rate can be simplified to $A_+\approx G^2\kappa_a/2J^2$ for $J/2\pi>0.1\,\mathrm{THz}$.
As a result, increasing the Q-factor of the WGM resonator as well as the photon-plasmon coupling strength $J$ can significantly reduce the light absorption $A_+$, thereby suppressing $\langle N_{\hat b}\rangle$. 
While increasing $G$ can heat molecular vibration mode, elevating $\langle N_{\hat b}\rangle$.

\begin{figure}
    \centering
    \includegraphics{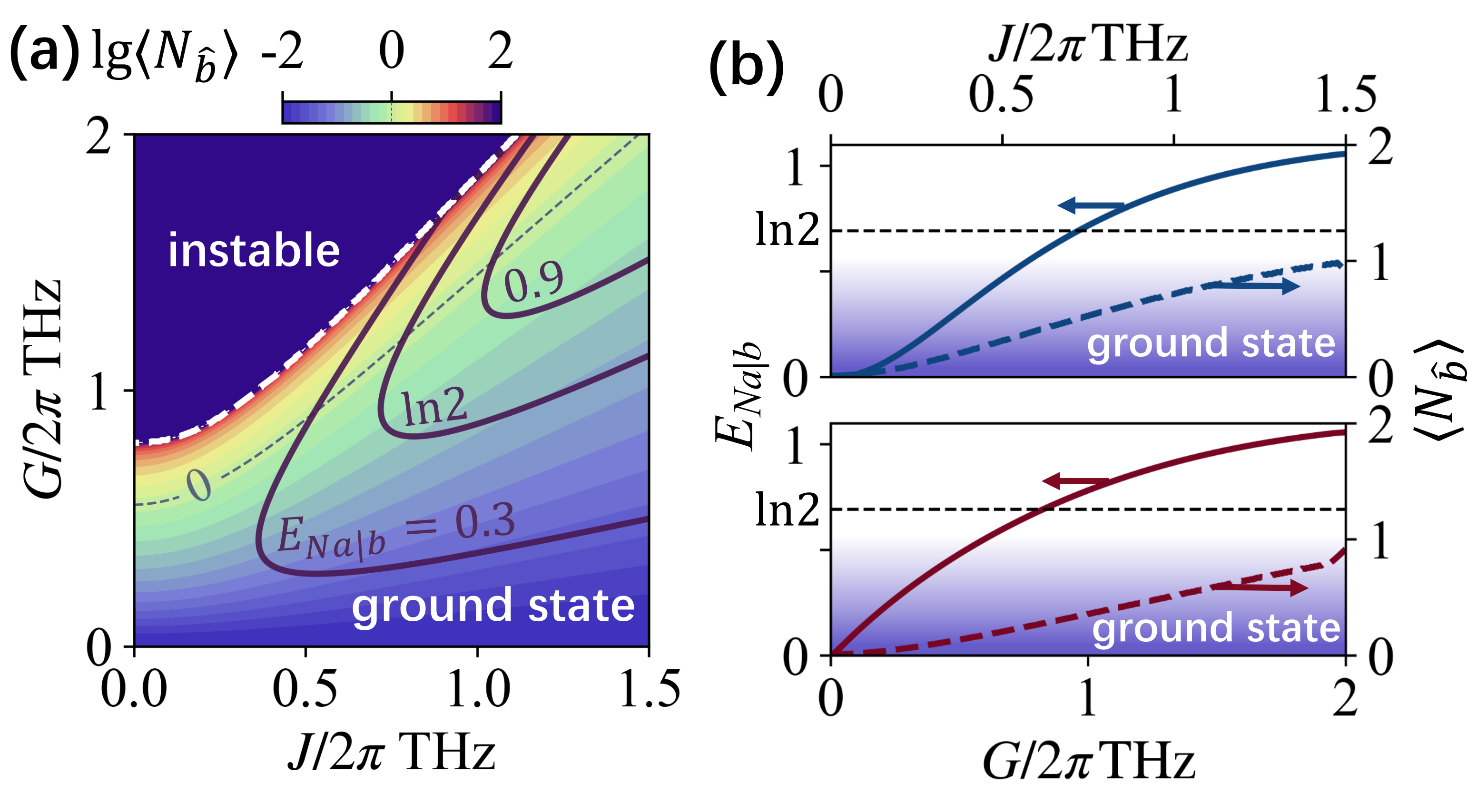}
    \caption{
    (a) Contour-filled plot of the stationary thermal excitation number $\langle N_{\hat b}\rangle$ with a contour line (dark blue dashed line) of $\langle N_{\hat b} \rangle=1$ ($\mathrm{lg}\langle N_{\hat b} \rangle=0$) and contour lines of logarithmic negativity $E_{Na|b}$ (dark purple solid line) versus $J$ and $G$.
    (b) Line plots of local maximum of $E_{Na|b}$ (solid line) and corresponding $\langle N_{\hat b}\rangle$ (dashed line) versus $J$($G$) by tuning $G$($J$). 
    We take $\Delta_c=\Delta_a=-\omega_b$, and the other parameters are the same as in Fig.\,\ref{fig:Fig2}.}
    \label{fig:Fig4}
\end{figure}

The stationary excitation number of phonon $\langle N_{\hat b}\rangle$ can also be directly derived from the CM $V$, with $\langle N_{\hat b}\rangle=(V_{33}+V_{44}-1)/2$ \cite{genesGroundstateCoolingMicromechanical}. Through numerical calculation, we plot it (color map) with $E_{Na|b}$ (contour lines) as a function of $J$ and $G$, as shown in Fig.\,\ref{fig:Fig4}(a).  Indeed,
$\langle N_{\hat b}\rangle$ decreases (increases) monotonically with the increase of $J$($G$),  aligning well with Eq.\,(\ref{eq:number}). 
Taken all together, the competition between $G$ and $J$ defines the entanglement value: larger $G$ amplifies two-mode squeezing (enhancing  $E_{Na|b}$) but increases thermal noise (suppressing $E_{Na|b}$), 
while stronger $J$ enhances photon redirection and suppresses thermal noise (both favoring $E_{Na|b}$) at the cost of reduced squeezing and Stokes photon storage (suppressing $E_{Na|b}$, see Eq.\,(\ref{eq:number}) and Eq.\,(S13a)). 
Therefore, only at the intermediate optimal $J$ and $G$ does the balance between photon redirection, squeezing, and noise suppression maximize entanglement.
Fig.\,\ref{fig:Fig4}(b) further presents the comparison of local maximal $E_{Na|b}$  and corresponding $\langle N_{\hat b}\rangle$ versus $J$ ($G$) by tuning $G$ ($J$). It can be seen that the optimal $E_{Na|b}$ under all parameters requires $\langle N_{\hat b}\rangle<1$, emphasizing ground state cooling as the critical factor to transcend conventional entanglement limits. 
One should note that the cooling process does not take energy away from the molecular vibration mode, but suppresses the absorption rate $A_+$. Thus the minimal achievable $\langle N_{\hat b}\rangle$ depends on the initial thermal phonon number $\bar n$,
and it is feasible for molecular COM systems to achieve ground state cooling with its $\bar n \approx 0.01$ at ambient temperature.

\prlsection{Conclusion}
In conclusion, we have presented a theoretical scheme to achieve robust photon-phonon entanglement under ambient conditions in a molecular COM system, which consists of an ultrahigh-Q WGM resonator coupled with a plasmonic nanocavity comprising a metallic nanoparticle and a single organic molecule \cite{maoWhisperinggalleryScanningMicroprobe2023,shlesingerIntegratedSidebandResolvedSERS2023}. 
This hybrid architecture allows the decoupling of entanglement storage (in the WGM resonator) from entanglement generation (localized at the plasmonic hotspot), enabling room-temperature quantum control in molecular systems—a regime previously considered challenging due to thermal noise. This capability is critical for applications such as quantum-enhanced sensing of molecular vibrations \cite{chikkaraddyMidinfraredperturbedMolecularVibrational2022,zhangEntanglementEnhancedSensingLossy2015} and on-chip quantum networks \cite{sunamiScalableNetworkingNeutralAtom2025,aghaeeradScalingNetworkingModular2025}, where scalability and noise tolerance are paramount.
In a broader view, we envision that future improvement of molecular optomechanical entanglement can be further implemented with more appealing quantum engineering techniques, such as dynamical modulation \cite{wangMacroscopicQuantumEntanglement2016,huManifestationClassicalNonlinear2019,huangQuadraturesqueezedLightOptomechanical2018}, reservoir engineering \cite{zhangEnhancedOptomechanicalEntanglement2020,wangReservoirEngineeredEntanglementOptomechanical2013,liaoReservoirengineeredEntanglementHybrid2018}, 
quantum interference \cite{tianRobustPhotonEntanglement2013}
and synthetic gauge fields \cite{liuPhasecontrolledAsymmetricOptomechanical2023}.
The hybrid platform compatibility with ambient conditions further positions it as a versatile tool for exploring macroscopic quantum phenomena in chemically rich environments.

\begin{acknowledgments}
The authors thank Jian Tang and Jun-Jie Li for helpful discussions.
T.J. is supported by the National Natural Science Foundation of China (NSFC, Grant No. 62471478).
H.J. is supported by NSFC (Grant No. 11935006, 12421005), the
National Key R\&D Program (2024YFE0102400), the Hunan Major Sci-Tech Program (2023ZJ1010).
Y.-F. J. is supported by the NSFC (Grant No. 12405029) and the Natural Science Foundation of Henan Province (Grant No. 252300421221).
K.W. is supported by the NSFC (Grant No. 12474351) and the Science Fund for Distinguished Young Scholars of Hunan Province (2024JJ2054).
Q.-R. L. is supported by Independent Innovation Science Foundation (24-ZZCX-BC-01).
\end{acknowledgments}

\nocite{*}

\bibliography{reference}

\end{document}


\makeatletter
\newcommand*{\addFileDependency}[1]{
  \typeout{(#1)}
  \@addtofilelist{#1}
  \IfFileExists{#1}{}{\typeout{No file #1.}}
}
\makeatother

\newcommand*{\myexternaldocument}[1]{%
    \externaldocument{#1}%
    \addFileDependency{#1.tex}%
    \addFileDependency{#1.aux}%
}
\myexternaldocument{main}

\renewcommand{\thefigure}{S\arabic{figure}}
\renewcommand{\thetable}{S\arabic{table}}
\renewcommand{\theequation}{S\arabic{equation}}


\preprint{APS/123-QED}

\title{Supplemental Information: Strong Molecule-Light Entanglement with Molecular Cavity Optomechanics}

\author{Hong-Yun Yu}
\affiliation{College of Advanced Interdisciplinary Studies, NUDT, Changsha 410073, P.R.China}
\author{Ya-Feng Jiao}
\affiliation{School of Electronics and Information, Zhengzhou University of Light Industry, Zhengzhou 450001, P.R.China}
\affiliation{Academy for Quantum Science and Technology, Zhengzhou University of Light Industry, Zhengzhou 450001, P.R.China}
\author{Jie Wang}
\affiliation{Key Laboratory of Low-Dimensional Quantum Structures and Quantum Control of Ministry of Education, Department of Physics and Synergetic Innovation Center for Quantum Effects and Applications, Hunan Normal University, Changsha 410081, P.R.China}

\author{Feng Li}
\affiliation{Key Laboratory for Physical Electronics and Devices of the Ministry of Education \& Shaanxi Key Laboratory of Information Photonic Technique, School of Electronic Science and Engineering, Faculty of Electronic and Information Engineering, Xi’an Jiaotong University, Xi’an 710049, P.R.China}
\author{Bin Yin}
\affiliation{Key Laboratory of Low-Dimensional Quantum Structures and Quantum Control of Ministry of Education, Department of Physics and Synergetic Innovation Center for Quantum Effects and Applications, Hunan Normal University, Changsha 410081, P.R.China}

\author{Tian Jiang}
\affiliation{Institute for Quantum Science and Technology, College of Science, NUDT, Changsha 410073, P.R.China}
\affiliation{College of Advanced Interdisciplinary Studies, NUDT, Changsha 410073, P.R.China}
\affiliation{Hunan Research Center of the Basic Discipline for Physical States,  Changsha 410073, P.R.China}

\author{Qi-Rui Liu}
\affiliation{College of Advanced Interdisciplinary Studies, NUDT, Changsha 410073, P.R.China}

\author{Hui Jing}
\affiliation{Key Laboratory of Low-Dimensional Quantum Structures and Quantum Control of Ministry of Education, Department of Physics and Synergetic Innovation Center for Quantum Effects and Applications, Hunan Normal University, Changsha 410081, P.R.China}
\affiliation{Institute for Quantum Science and Technology, College of Science, NUDT, Changsha 410073, P.R.China}

\author{Ke Wei}
\affiliation{Institute for Quantum Science and Technology, College of Science, NUDT, Changsha 410073, P.R.China}
\maketitle

This Supplementary Information is organized as follows: In Sec,\,\ref{sec:Hamiltonian}, we elucidate the derivation of  quantum Langevin
equations and the methods of calculating logarithmic negativity $E_N$ from covariance matrix (CM) $V$. In Sec.\,\ref{sec:stokes photon} we derive the response function of plasmonic mode and optical mode toward Stokes photon. In Sec.\,\ref{sec:purterbation}, we derive the expression for the absorption $A_+$ and emission $A_-$ rate of molecular vibration mode. In Sec.\,\ref{sec:single mode}, we illustrates the feasibility of single mode operation in WGM resonator. In Sec\,\ref{sec:N}, we proves the validity of illustrating effective stationary excitation number with plasmonic power spectrum. In Sec.\,\ref{sec:Nexperi} we proposes a practicable experiment setup for entanglement validation.

\tableofcontents 
\section{\label{sec:Hamiltonian} Quantum Langevin equations}
The dynamical equations for the three-mode annihilation operators $\hat a,\hat b,\hat c$ are derived by applying the quantum Langevin equations,
expressed as:
\begin{subequations}
\begin{eqnarray}
&&\dot{\hat{a}}=-(i\Delta_a+\kappa_a)\hat{a}-iJ\hat c+\sqrt{2\kappa_a}\hat a_{\mathrm{in}},\\
&&\dot{\hat{b}}=-(i\omega_b+\gamma)\hat{b}+ig_c\hat c^\dagger \hat c+\sqrt{2\gamma}\hat b_{\mathrm{in}},\\
&&\dot{\hat c}=-(i\Delta_c+\kappa_c)\hat c+ig_c\hat c(\hat{b}^\dagger+\hat{b})-iJ\hat a+\epsilon_L+\sqrt{2\kappa_c}\hat c_{\mathrm{in}},
\end{eqnarray}
\label{eq:QLE}
\end{subequations}
where $\hat a_{\mathrm{in}}$ and $\hat c_{\mathrm{in}}$ are Gaussian input noise operators for the optical and plasmonic modes, respectively, describing the zero-point fluctuations in the fiber-guided optical mode and in the vacuum environment. The Langevin force operator $\hat b_{\mathrm{in}}$ accounts for the Brownian motion of the molecular vibrational mode. These noise operators obey the following quantum correlation relations:
\begin{subequations}
\begin{eqnarray}
&&\langle \hat a_{\mathrm{in}}(t)\hat a_{\mathrm{in}}^\dagger(t^{\prime})\rangle=\delta(t-t^{\prime}),\\
&&\langle \hat b_{\mathrm{in}}(t)\hat b_{\mathrm{in}}^\dagger(t^{\prime})\rangle=(\bar n+1)\delta(t-t^{\prime}),\\
&&\langle \hat b_{\mathrm{in}}^\dagger(t)\hat b_{\mathrm{in}}(t^{\prime})\rangle=\bar n\delta(t-t^{\prime}),\\
&&\langle \hat c_{\mathrm{in}}(t)\hat c_{\mathrm{in}}^\dagger(t^{\prime})\rangle=\delta(t-t^{\prime}),
\end{eqnarray}
\end{subequations}
where a Markov approximation has been made, which is valid for a large mechanical quality factor $Q=\omega_b/\gamma\gg1$. $\bar{n}=1/[\exp(\omega_{b}/k_{B}T)-1]$ is the mean thermal occupation number of the single molecular vibration mode for environmental temperature $T$, with $k_{B}$ the Boltzmann constant. Since the typical frequency of molecular vibration is $\omega_b\sim O(10)\,\mathrm{THz}$, the thermal occupation number can be rigorously approximated as $\bar n\approx 0.01$ under ambient conditions.

Under the strong driving condition of the plasmonic mode,
we linearize the system dynamics in Eq.\,(\ref{eq:QLE}) by expanding each operator as a sum of its steady-state mean value and a quantum fluctuation around it: $\hat{a}=a_s+\delta \hat{a}, \hat{b}=b_s+\delta \hat{b}, \hat c=c_s+\delta \hat c$, where the steady-state mean values $a_s$, $b_s$ and $c_s$ are obtained as
\begin{equation}
    a_s=\frac{-iJc_s}{i\Delta_a+\kappa_a},b_s=\frac{ig_c|c_s|^2}{i\omega_b+\gamma},c_s=\frac{-iJa_s+\epsilon_L}{i\Delta_s+\kappa_c},
\end{equation}
where $\Delta_s=\Delta_c-g_c(b_s^\dagger+b_s)$.
Our results in the main text reveals dominant quantum effects near the critical detuning conditions $\Delta_c\approx \pm \omega_b$. This resonance configuration permits the adoption of the parameter equivalence $\Delta_s=\Delta_c$ with good approximation. Throughout this work, we therefore maintain the simplified detuning relationship $\Delta_s=\Delta_c$ to streamline theoretical developments.
Correspondingly, by neglecting the second-order fluctuation terms, the linearized dynamical equations for fluctuation are obtained as
\begin{subequations}
\label{eq:fluctuation}
\begin{align}
&\dot{{\delta}{\hat{a}}} =-(i\Delta_a+\kappa_a)\delta \hat{a}-iJ\delta \hat c+\sqrt{2\kappa_a}\hat a_{\mathrm{in}},\\
&\dot{{\delta}{\hat{b}}} =-(i\omega_b+\gamma)\delta \hat{b}+ig_c(c_s^\dagger\delta \hat c+c_s\delta \hat c^\dagger)+\sqrt{2\gamma}\hat b_{\mathrm{in}},\\
&\dot{{\delta}{\hat c}} =-(i\Delta_c+\kappa_c)\delta\hat  c+ig_cc_s(\delta \hat{b}^\dagger+\delta \hat{b}) -iJ\delta \hat{a}+\sqrt{2\kappa_c}\hat c_{\mathrm{in}}.
\end{align}
\end{subequations}
By defining the operators $\hat X_o=(\delta \hat o^\dagger+\delta \hat o)/\sqrt{2},\quad \hat Y_o=i(\delta \hat o^\dagger-\delta \hat o)/\sqrt{2}$ and $\hat X_{o_{\mathrm{in}}}=(\hat o_{\mathrm{in}}^\dagger+\hat o_{\mathrm{in}})/\sqrt{2},\quad \quad \hat Y_{o_{\mathrm{in}}}=i(\hat o_{\mathrm{in}}^\dagger- \hat o_{\mathrm{in}})/\sqrt{2}$ for
 $o \in\{{a},{b}, c\}$, the linearized dynamical equation for quadrature fluctuations can be expressed as
\begin{align}
\dot{\hat{u}}(t)=A u(t)+n(t),
\end{align}
where $u(t)$ and $n(t)$ are the vectors of quadrature fluctuations and input noises, expressed as:
\begin{align}
 & \hat u(t)=(\hat X_a,\hat Y_a,\hat X_b,\hat Y_b,\hat X_c,\hat Y_c)^T, \\ &\hat n(t)=\sqrt{2}(\sqrt{\kappa_a}\hat X_{a_\mathrm{in}},\sqrt{\kappa_a}\hat Y_{a_\mathrm{in}},\sqrt{\gamma}\hat X_{b_\mathrm{in}},\sqrt{\gamma}\hat Y_{b_\mathrm{in}},\sqrt{\kappa_c}\hat X_{c_\mathrm{in}},\sqrt{\kappa_c}\hat Y_{c_\mathrm{in}})^T,
\end{align}
The drift matrix $A$ is given by:
 \begin{equation}
 \label{eq:matrixA}
     A=
     \begin{pmatrix}
     -\kappa_a&\Delta_a&0&0&0&J\\
     -\Delta_a&-\kappa_a&0&0&-J&0\\
     0&0&-\gamma&\omega_b&0&0\\
     0&0&-\omega_b&-\gamma&G &0\\
     0&{J}&0&0&-\kappa_c&\Delta_c\\
     -{J}&0&G&0&-\Delta_c&-\kappa_c
     \end{pmatrix},
 \end{equation}
where the effective plasmon-phonon coupling strength $G=2g_cc_s$ is assumed to be real, which can be achieved by choosing suitable phase reference of the optical field. The linearized dynamics exhibit asymptotic stability if and only if all eigenvalues $\lambda_i$ of the drift matrix $A$ satisfy $\mathrm{Re}(\lambda_i)<0$ \cite{SstrogatzNonlinearDynamicsChaos2015,SdejesusRouthHurwitzCriterionExamination1987}.

Within this stable regime, due to the linearized dynamics and the Gaussian nature of quantum noises, the steady state of the quantum fluctuations of this hybrid COM system can finally evolve into a zero mean three-mode Gaussian state, which can be completely characterized by an $6\times6$ covariance matrix (CM) $V$ with its entries defined as $$V_{ij}=\langle u_i(t)u_j(t')+ u_j(t')u_i(t)\rangle/2~(i,j=1,2,...,6).$$ The steady-state CM $V$ is governed by the Lyapunov equation:
 \begin{equation}
\label{eq:calculate_V}
AV+VA^T=-D,
\end{equation}
where $D=\mathrm{diag}[\kappa_a,\kappa_a,\gamma(2 \bar n+1),\gamma(2 \bar n+1),\kappa_c,\kappa_c]$ is the diffusion matrix, and it is defined through $D_{kl}\delta (s-s')=\langle n_{k}(s)n_{l}(s')+n_{l}(s')n_{k}(s)\rangle/2$. The Lyapunov equation (\ref{eq:calculate_V}) is linear for $V$ and can be solved straightforwardly, but its general exact expression are complex and will not be reported here.
To explore the behavior of bipartite entanglement in this multimode continuous variable system, we employ the quantitative entanglement measure logarithmic negativity $E_N$, defined as \cite{SadessoExtremalEntanglementMixedness2004}:
\begin{align}
E_N=\mathrm{max}\lbrack 0,-\ln(2v^-_{m,n})\rbrack, \label{En}
\end{align}
where $v^-_{m,n}=\{\sum(V_{mn})-\lbrack\sum(V_{mn})^2-4\mathrm{det}V_{mn}\rbrack^{1/2}\}^{1/2}/\sqrt{2}$, with $\sum{(V_{mn})}=\mathrm{det}\mathcal{A}+\mathrm{det}\mathcal{B}-2\mathrm{det}\mathcal{C}$. $v^-_{m,n}$ is the minimum symplectic eigenvalue of the partial transpose of a reduced $4\times 4$ CM $V_{mn}$, with $m$ and $n$ describing the two selected
modes under consideration. The reduced CM $V_{mn}$ is extracted from the full CM $V$ by selecting the rows and columns with respect to the selected bipartition $m$ and $n$, which can be rewritten in a $2\times 2$ block form as
\begin{equation}
    V_{mn}=
    \begin{pmatrix}
    \mathcal{A}&\mathcal{C}\\
    \mathcal{C}^T&\mathcal{B}
    \end{pmatrix}.
\end{equation}
Eq.\,(\ref{En}) indicates that the selected bipartition $m$ and $n$ gets entangled if and only if $v^{-}_{m,n}<1/2$,
where $E_N$ has a nonzero value. It should be emphasized that $E_N$ quantifies how much the positivity of the partial transpose condition for separability is violated for the considered Gaussian states, which is equivalent to Simon's necessary and sufficient entanglement nonpositive partial transpose criterion (or the related Peres-Horodecki criterion).

In addition, we illustrate the feasibility of the range of the effective plasmon-phonon coupling strength $G$. the single-photon coupling strength $g_{c}$ in molecular COM systems is typically $10-10^{2}\,\mathrm{GHz}$ \cite{SroelliMolecularCavityOptomechanics2016}. Given that the plasmonic nanocavity cannot tolerate intense pumping ($\sim O(10)$\,mW) \cite{SsunRevealingPhotothermalBehavior2022}, the corresponding effective coupling strength $G$ will still be limited to the weak coupling regime. For example, we have confirmed that the steady-state mean value of the plasmonic mode is about $|c_s|\approx10$, with $P=8\,\mathrm{mW}$, $\Delta_c=\Delta_a=-\omega_b,J=0$, and $g_c/2\pi=20 \,\mathrm{GHz}$, which corresponds to an effective coupling strength of $G/2\pi\approx0.4\,\mathrm{THz}$. In this case, we consider $G/2\pi$ is within the range $[0,2]\,\mathrm{THz}$, where the values of $G$ can be modulated by adjusting the pump power or positioning the molecule at different locations within the plasmonic nanocavity. It is also noted that we consider $J/2\pi$ is within the interval of $[0,1.5]\,\mathrm{THz}$, where the values of $J$ can be tuned by positioning metallic nanoparticles at different locations within the evanescent field of WGM resonator~\cite{SshlesingerHybridCavityantennaArchitecture2023,SshlesingerIntegratedMolecularOptomechanics2021}.

Furthermore, in previous studies on stationary entanglement where pump fields resonantly couple to Stokes/anti-Stokes sidebands, the rotating wave approximation (RWA) has been conventionally employed under the good-cavity limit to derive analytical expressions for logarithmic negativity \cite{SgenesRobustEntanglementMicromechanical2008,SwangReservoirEngineeredEntanglementOptomechanical2013}. However, this paradigm becomes inadequate for molecular optomechanical systems incorporating plasmonic nanocavities, where the inherent high plasmonic loss violate the good-cavity condition and significantly degrade the accuracy of RWA. Therefore, Section IV of our main text presents phenomenological explanations. Specifically, our framework provides physical insights into entanglement optimization between the coupling strengths $J$ and $G$, though precise analytical determination of their optimal conditions remains intractable in molecular optomechanical systems with large plasmonic loss.

\section{\label{sec:stokes photon} Stokes photon response amplitude}
The response amplitudes of the optical ($f_a[-\omega_b]$) and plasmonic ($f_c[-\omega_b]$) modes to Stokes photons are derived by analyzing the linearized fluctuation dynamics under a virtual $-\omega_b$ driving term. This is equivalent to replacing the optomechanical coupling term  $ig_cc_s(\delta \hat b^\dagger+\delta \hat b)$ in Eq. (S4c) with a unit driving term $e^{i\omega_b t}$. Transforming Eq.\,(\ref{eq:fluctuation}a) and Eq.\,(\ref{eq:fluctuation}c) to the frequency domain yields:  
\begin{subequations}
\begin{align}
    i\omega_b f_c[-\omega_b]&=-(i\Delta_c+\kappa_c)f_c[-\omega_b]-iJf_a[-\omega_b]+1,\\
    i\omega_b f_a[-\omega_b]&=-(i\Delta_a+\kappa_a)f_a[-\omega_b]-iJf_c[-\omega_b].
\end{align}
\end{subequations}
From which we can obtain:
\begin{subequations}
\begin{align}
    f_a[-\omega_b]&=-\frac{iJ}{J^2+[\kappa_a+i(\omega_b+\Delta_a)][\kappa_c+i(\omega_b+\Delta_c)]}, \label{eq:response-a}\\
    f_c[-\omega_b]&=\frac{\kappa_a+i(\omega_b+\Delta_a)}{J^2+[\kappa_a+i(\omega_b+\Delta_a)][\kappa_c+i(\omega_b+\Delta_c)])}.
    \label{eq:response-c}
\end{align}
\label{eq:Fano}
\end{subequations}
Thus, the ratio of optical-to-plasmonic response amplitudes is given by:
\begin{equation}
    R_{ac}\equiv\frac{f_a[-\omega_b]}{f_c[-\omega_b]}=-\frac{iJ}{\kappa_a+i(\omega_b+\Delta_a)}.
    \label{eq:Sratio}
\end{equation}
From Eq.\,(\ref{eq:response-a}) and Eq.\,(\ref{eq:Sratio}), the increase in $J$ not only enhances $R_{ac}$ --- signifying improved Stokes photon redirection (thereby amplifying entanglement) ---
but also reduces $f_a[-\omega_a]$, which weakens the efficiency of Stokes photon storage (thereby suppressing entanglement).
Consequently, both excessively low and high values of $J$ limit entanglement. Only at an intermediate optimal $J$ does the trade-off among photon redirection, photon storage and two-mode squeezing maximize entanglement, as rigorously demonstrated in Fig.\,\ref{fig:Fig3} and Fig.\,\ref{fig:Fig4}(a) in the main text.

\section{\label{sec:purterbation} Optomechanical perturbation theory}
In the weak coupling regime ($G\ll \omega_b,\kappa_c$), optomechanical interaction can be analyzed with perturbation theory, where optomechanical coupling is assumed to be the perturbation of the plasmonic field. The molecular vibration mode experiences two noise baths: the thermal bath, characterized by damping rate $\gamma$ and thermal occupation $\bar n$, and radiation-pressure noise arising from fluctuations of the plasmonic mode, quantified by the noise power spectrum $S_{FF}(\omega)$, obtaining through Fourier transformation of the autocorrelation function of radiation-pressure force $F=G(\delta \hat{c}+\delta \hat{c}^\dagger)/2x_{\mathrm{zpf}}$ \cite{SliuReviewCavityOptomechanical2013}:
\begin{align}
     S_{FF}(\omega)&=\int \langle F(t)F(0)\rangle e^{i\omega t} dt\nonumber \\&= \int \langle F^\dagger(\omega)F(\omega')\rangle d\omega' \nonumber,
\end{align}
where $x_{\mathrm{zpf}}=\sqrt{\hbar/2m\omega_b}$ is the zero-point fluctuation amplitude of the molecular vibration.  
The photon absorption ($A_+$) and emission ($A_-$) rates are then given by: $A_\pm=S_{FF}(\mp\omega_b) x_{\mathrm{zpf}}^2$, and finally the effective stationary excitation numbers of phonon is calculated via \cite{SaspelmeyerCavityOptomechanics2014,SliuReviewCavityOptomechanical2013}:
\begin{equation}
    \langle N_{\hat b} \rangle=\frac{2\bar n \gamma +A_+}{2\gamma-A_++A_-}
    \label{eq:Snumber}.
\end{equation}
To calculate $S_{FF}(\omega)$, we neglect optomechanical back-action of molecule on plasmonic field (valid for $G\ll \kappa_c$, the term $ig_cc_s(\delta \hat{b}^*+\delta \hat{b})$ in Eq.\,(\ref{eq:fluctuation})), the noise power spectrum is derived from the linearized equations\,(\ref{eq:fluctuation}a-c) in the frequency domain. Under resonance $\Delta_c=-\omega_b$:
\begin{align}
S_{FF}(\omega&,J,\Delta_a)=\nonumber\\
&\frac{G^2}{2x_{\mathrm{zpf}}^2}\left(J^2 \kappa_a+\kappa_c \left(\kappa_a^2+(\omega-\Delta_a)^2\right)\right)/\nonumber\\
&\Bigl ( J^4+2 J^2 (\kappa_a \kappa_c-(\omega-\Delta_a) (\omega+\omega_b))+\nonumber\\
&\left(\kappa_a^2+(\omega-\Delta_a)^2\right) \left(\kappa_c^2+(\omega+\omega_b)^2\right) \Bigr) \label{eq:SFF}.
\end{align}
The absorption rate is \cite{SclerkIntroductionQuantumNoise2010}:
\begin{align}
    A_+=&S_{FF}(-\omega_b)x_{\mathrm{zpf}}^2=\nonumber\\
    &\frac{G^2}{2}\frac{ \kappa_a(J^2+\kappa_a\kappa_c)+\kappa_c(\omega_b+\Delta_a)^2}{(J^2+\kappa_a\kappa_c)^2+\kappa_c^2(\omega_b+\Delta_a)^2}
    \label{eq:Aplus0}.
\end{align}
Under $\Delta_a=-\omega_b$, $A_+$ simplifies to:
\begin{align}
    A_+=\frac{G^2}{2}\frac{ \kappa_a}{J^2+\kappa_a\kappa_c} \label{eq:Aplus}.
\end{align}
Following the same steps, we obtain the emission rate ($\kappa_a\ll J\ll \omega_b,\kappa_c$):
\begin{equation}
    A_-=\frac{G^2}{2}\frac{J^2\kappa_a+(4\omega_b^2+\kappa_a^2)\kappa_c}{J^4+J^2(-8\omega_b^2+2\kappa_a\kappa_c)+(4\omega_b^2+\kappa_a^2)(4\omega_b^2+\kappa_c^2)}\approx\frac{G^2}{2}\frac{\kappa_c}{4\omega_b^2+\kappa_c^2},
\end{equation}
which is insensitive with WGM resonator's coupling.

When $\Delta_a=-\omega_b$ and $\kappa_a\ll\gamma$, the absorption rate $A_+$ 
must account for the finite linewidth of the molecular vibration mode:
\begin{equation}
    A_+=\int_{-\infty}^0 d\omega S_{FF}(\omega) \frac{\gamma/\pi}{(\omega+\omega_b)^2+\gamma^2}
    \label{eq:modified Aplus},
\end{equation}
where $(\gamma/\pi)/\left((\omega+\omega_b)^2+\gamma^2\right)$ represents the normalised response function of molecular vibration mode. This integral incorporates the Lorentzian response of the molecular mode, ensuring consistency between $\langle N_{\hat b}\rangle$ calculated from the covariance matrix $V$ and Eq.\,(\ref{eq:Snumber}), which will be further discussed in Sec.\ref{sec:N}.

\section{\label{sec:single mode}Necessity of Single-Mode Operation for Photon-Phonon Entanglement}
\begin{figure}
    \centering
    \includegraphics[width=0.6\textwidth]{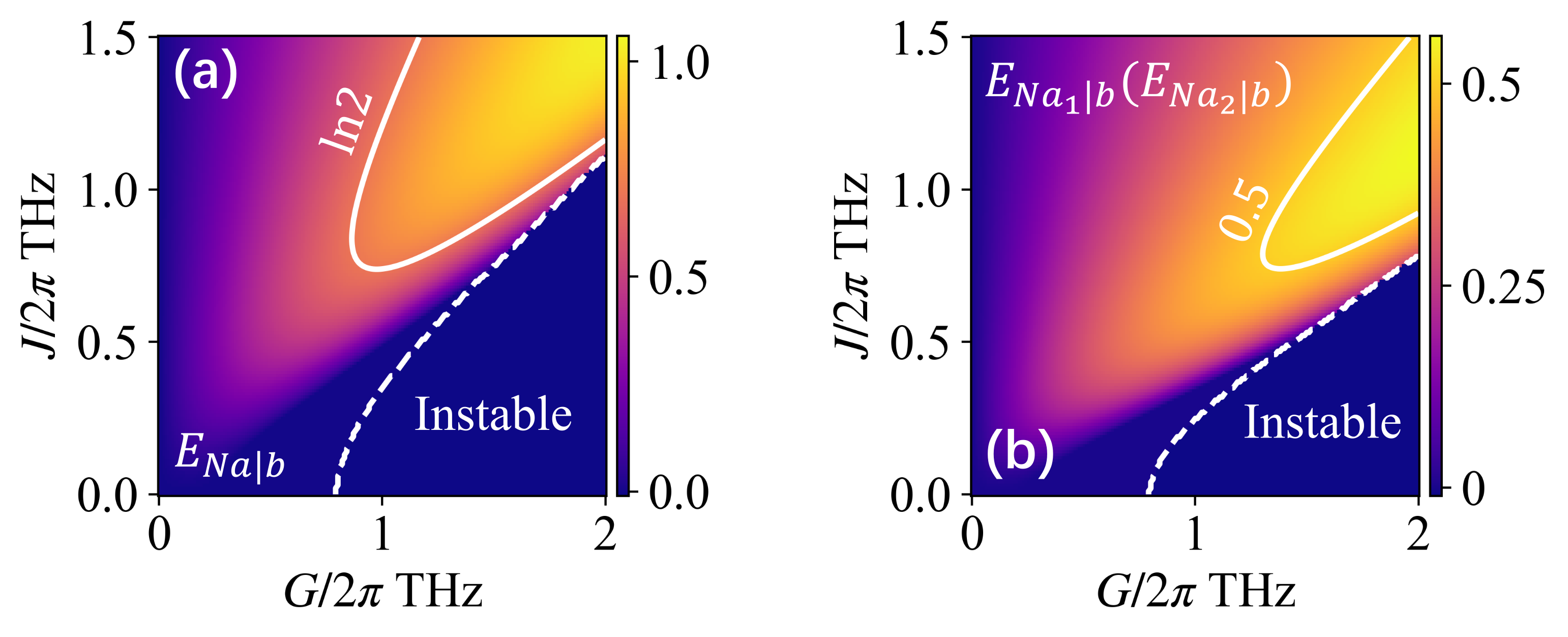}
    \caption{Density plots of $E_{Na|b}$ versus $J$ and $G$ with (a) single mode (b) two degenerate modes. We take $\Delta_a/\omega_b=\Delta_c/\omega_b=-1,\kappa_a/2\pi=10^{-3}\,\mathrm{THz}$. The other parameters are the same with Fig.\,\ref{fig:Fig2} in the main text.}
    \label{fig:Fig6}
\end{figure}
While the WGM resonator can support both clockwise (CW) mode and counter-clockwise (CCW) mode, we only consider the CW mode coupling to the plasmon in the main text. Typically, this single-mode operation is achieved by strategically placing two scattering nanoparticles near the WGM field \cite{SsvelaCoherentSuppressionBackscattering2020,SwiersigStructureWhisperinggalleryModes2011}, which induces asymmetric backscattering between the two propagation modes. In our proposed configuration, although only one real metal nanoparticle is placed, the backscattering of the cavity itself produces a phase difference \cite{SsvelaCoherentSuppressionBackscattering2020}, which can be regarded as a virtual particle, resulting in asymmetric backscattering in the microdisk cavity and guarantying single-mode operation.

More importantly, numerical simulations confirm that single-mode operation is not strictly necessary for photon-phonon entanglement ($E_{Na|b}$) but optimizes it. When both CW and CCW modes coexist (Fig.\,\ref{fig:Fig6}), the molecular vibration mode entangles with both optical modes, reducing $E_{Na|b}$ by $\sim50\%$. This reduction can also be mitigated by  asymmetric backscattering or lifting mode degeneracy through external tuning, as demonstrated in previous research on nonreciprocal optomechanical systems~\cite{SjiaoNonreciprocalOptomechanicalEntanglement2020}.

\section{\label{sec:N} Validity of Thermal Noise Calculating via $S_{FF}(\omega)$}
The calculation of the molecular vibration mode’s thermal occupation $\langle N_{\hat b}\rangle$ from the noise power spectrum  $S_{FF}(\omega)$ requires careful consideration of the interplay between the plasmonic field’s noise characteristics and the molecular mode’s response. Here, we rigorously validate this approach and address corrections necessary for systems where the WGM linewidth $\kappa_a$  is smaller than the molecular damping rate $\gamma$.
\subsection{Theoretical Framework}
The standard expression for the absorption rate $A_+$ 
requires
$\gamma\ll\kappa_a,\kappa_c$ \cite{SmarquardtQuantumTheoryCavityAssisted2007}, so that molecular vibration responds only at frequencies $\pm \omega_b$:
\begin{equation}
    A_+=S_{FF}(-\omega_b)x_{\mathrm{zpf}}^2,
\end{equation}
where $x_{\mathrm{zpf}}=\sqrt{\hbar/2m\omega_b}$ is the zero-point fluctuation amplitude. This approximation holds when the noise spectrum $S_{FF}(\omega)$ is flat near $-\omega_b$, i.e., $\kappa_a\gg\gamma$.   
However, for high-Q WGM resonators ($\kappa_a/2\pi=10^{-4}\,\mathrm{THz}\ll\gamma/2\pi=10^{-2}\,\mathrm{THz}$), $S_{FF}(\omega)$ exhibits sharp spectral features (Fig.\,\ref{fig:FigS2}b), comparable with 
the molecular mode’s Lorentzian response bandwidth $\gamma$, necessitating integration over the negative frequency range.  
We now deduce the corrected absorption rate. As the molecular vibration mode acts as a harmonic oscillator with susceptibility:
\begin{equation}
    \chi(\omega)=\frac{1}{\omega_b^2-\omega^2-2i\gamma\omega}.
\end{equation}
The power absorbed from the plasmonic noise is proportional to the overlap between $S_{FF}(\omega)$ and $\mathrm{Im}[\chi(\omega)]$ on the negative frequency, which peaks near $\omega=-\omega_b$, with a width $\gamma$. From Fermi's golden rule, we have the transition rate $A_+\propto\int_{-\infty}^0 S_{FF}(\omega)\mathrm{Im}[\chi(\omega)]d\omega$. 
For $\omega_b\gg\gamma$ and near $\omega=-\omega_b$, we have $\mathrm{Im}[\chi(\omega)]\approx\frac{1}{2\omega_b}\frac{\gamma}{(\omega+\omega_b)^2+\gamma^2}$. Thus, the corrected absorption rate becomes:  
\begin{equation}
    A_+=\int_{-\infty}^0  S_{FF}(\omega)\cdot \mathcal{L}(\omega)d\omega=\int^0_{-\infty}S_{FF}(\omega) \frac{\gamma/\pi}{(\omega+\omega_b)^2+\gamma^2}d\omega,
    \label{eq:mAplus}
\end{equation}
where $\mathcal{L}(\omega)=(\gamma/\pi)/\left((\omega+\omega_b)^2+\gamma^2\right)$ is the molecular mode’s Lorentzian response (Fig.\,\ref{fig:FigS2}(c)). Eq.\,(\ref{eq:mAplus}) accounts for the finite linewidth $\gamma$, ensuring consistency with the covariance matrix (CM) formalism.

\subsection{Numerical Validation}

We compare $\langle N_{\hat b}\rangle$ calculated via three methods: Direct solution of $AV+VA^T=-D$, conventional $A_+=S_{FF}(-\omega_b)x_{\mathrm{zpf}}^2$, and modified from $A_+$ from Eq.\,(\ref{eq:mAplus}), denoting as $\langle N_{\hat b}\rangle_{CM}$,$\langle N_{\hat b}\rangle_{S}$ and $\langle N_{\hat b}\rangle_{S}'$, respectively.

\begin{figure}
    \centering
    \includegraphics[width=0.6\textwidth]{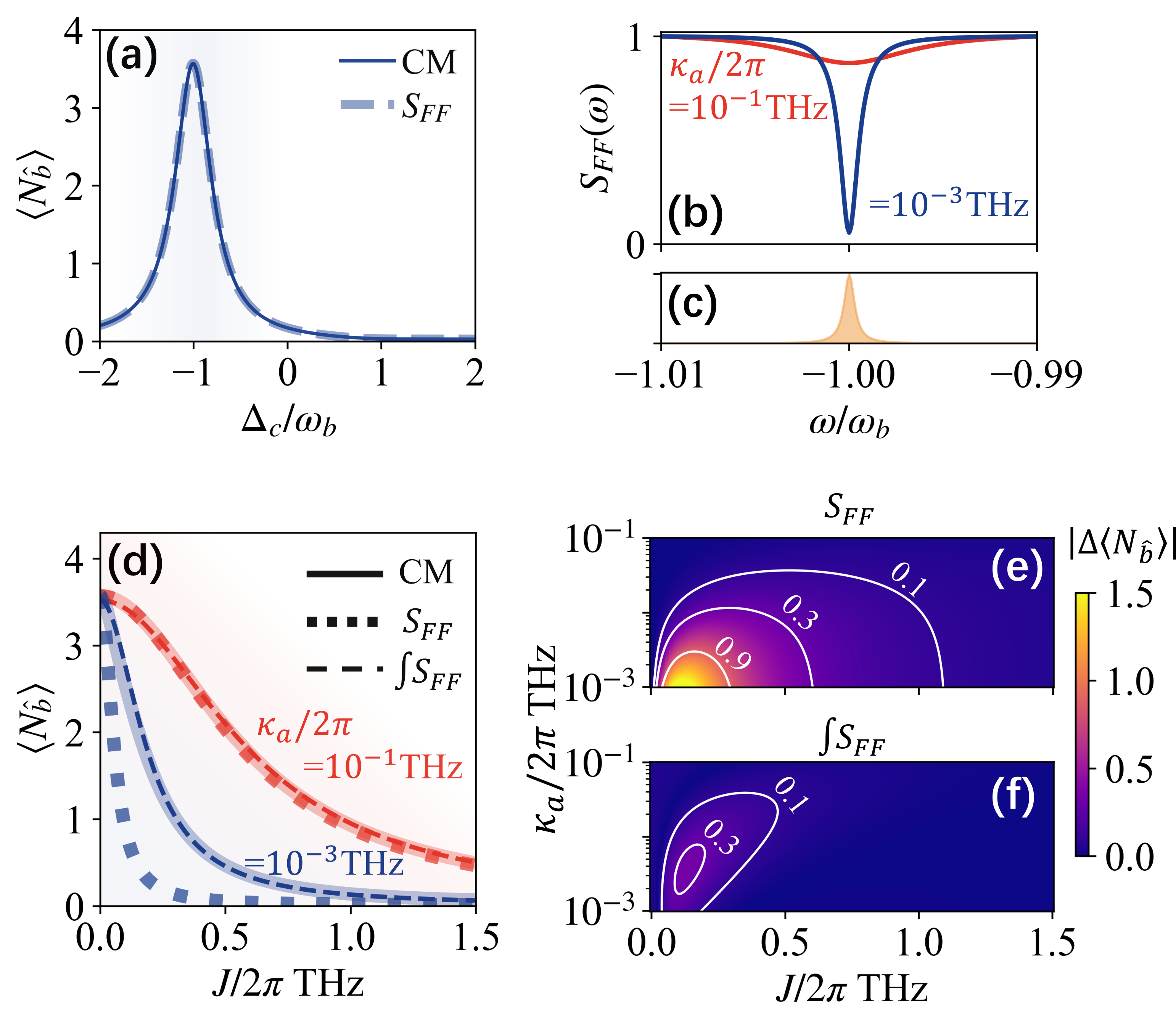}
    \caption{(a) $\langle N_{\hat b}\rangle$ derived from CM (solid line) and $S_{FF}(\omega)$ (dashed line) versus $\Delta_c$. (b) Noise power spectrum $S_{FF}(\omega)$ around the frequency of $-\omega_b$ for different values of $\kappa_a$. (c) Normalised response function of molecular vibration mode. (d) $\langle N_{\hat b}\rangle$ derived from CM, $S_{FF}(\omega)$, and modified $S_{FF}(\omega)$ versus $J$ for different values of $\kappa_a$. Density Plots of $\langle N_{\hat b}\rangle_{CM}$ absolute difference($|\Delta\langle N_{\hat b}\rangle|$) with (e)$\langle N_{\hat b}\rangle_{S}$ and (f)$\langle N_{\hat b}\rangle_{S}'$ versus $J$ and $\kappa_a$. We take $J=0$ in (a), $J/2\pi=0.5\,\mathrm{THz}$ in (b), $G/2\pi=0.7\,\mathrm{THz}$ and $\Delta_a=\Delta_c=-\omega_b$ in (a)-(f).}
    \label{fig:FigS2}
\end{figure}

Firstly, the case of $\kappa_a\gg\gamma$ is discussed, this is valid for $J=0$ (that is, no photon-plasmon coupling and thus $\kappa_a$ is not considering) or $J\neq 0,\kappa_a/2\pi \gg \gamma/2\pi=0.01\,\mathrm{THz}$. 
As shown in Fig.\,\ref{fig:FigS2}(a) ($J=0$) and the red lines in Fig.\,\ref{fig:FigS2}(d) ($J/2\pi=0.5\,\mathrm{THz}$, $\kappa_a/2\pi=0.1\,\mathrm{THz}$),  
$\langle N_{\hat b}\rangle_{CM}$ and  $\langle N_{\hat b}\rangle_{S}$ agree perfectly, 
as $S_{FF}(\omega)$ is flat near $-\omega_b$ (Fig.\,\ref{fig:FigS2}(b), red line), and thus the Markov approximation holds, validating $A_+=S_{FF}(-\omega_b)x_{\mathrm{zpf}}^2$.

Then we consider the case of $\kappa_a\ll\gamma$, as shown in Fig.\,\ref{fig:FigS2}(d), blue lines, for $J/2\pi=0.5\,\mathrm{THz}$ and $\kappa_a/2\pi=10^{-4}\,\mathrm{THz}$,
$\langle N_{\hat b}\rangle_{S}$ underestimates $\langle N_{\hat b}\rangle_{CM}$ due to spectral narrowing, 
while $\langle N_{\hat b}\rangle_{S}'$ restores agreement with CM results.

For a general case, as shown in Figs.\,\ref{fig:FigS2}(e,f), Density plots of $|\Delta\langle N_{\hat b}\rangle|=|\langle N_{\hat b}\rangle_{CM}-\langle N_{\hat b}\rangle_S|$ show significant discrepancies for $\kappa_a\ll\gamma$, which can be resolved well by the modified $A_+$, confirming our analysis of thermal phonon number $\langle N_{\hat b}\rangle$ with $S_{FF}(\omega)$.

\section{\label{sec:Nexperi}Experimental Protocol for Entanglement Verification}

The Raman scattering enhancement effects in hybrid plasmonic systems with high-Q cavity are not universally guaranteed, as demonstrated by multiple experimental studies \cite{SshlesingerHybridCavityantennaArchitecture2023,SshlesingerIntegratedSidebandResolvedSERS2023b}.
Raman enhancement factors depend critically on two key parameters: (i) pump field enhancement and (ii) the local density of states (LDOS) enhancement at the Stokes sideband. 
These parameters can be effectively modulated  through coupling with high-Q cavity.
Raman enhancement can be either significantly suppressed or enhanced when the pump frequency or Stokes sideband aligns with Fano resonance features (dip/peak) \cite{SshlesingerHybridCavityantennaArchitecture2023,SshlesingerIntegratedSidebandResolvedSERS2023b}.

\begin{figure}
    \centering
    \includegraphics{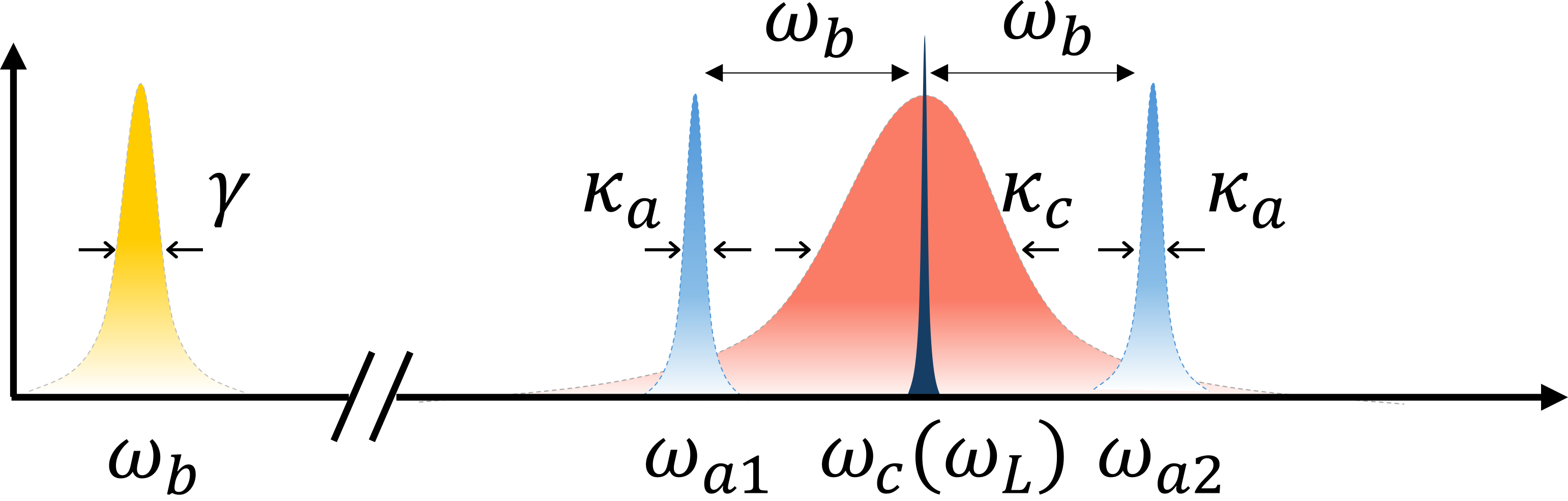}
    \caption{Frequency spectrum of the system with two WGMs}
    \label{fig:FigS4}
\end{figure}

In the main text, we employ WGM resonator coupling to control the LDOS, which governs the molecular light absorption rate $A_+$.
The pump enhancement is quantified through the effective plasmon-phonon coupling strength $G=2g_cc_s$.
in which the influence of coupling WGM resonators on single-photon coupling strength $g_c$ has not been taken into account.
Moreover, practical implementation faces two challenges: (1) the required pump detuning ($\Delta_c = -\omega_b$) makes plasmonic nanocavity excitation experimentally demanding, and (2) direct measurement of terahertz-frequency molecular vibrations presents significant technical hurdles.
To overcome these limitations, we introduce a second WGM resonator resonant at $\Delta_{a2} = \omega_b$ with plasmon-photon coupling strength $J_2$ (Fig. \ref{fig:FigS4}). This configuration redirects anti-Stokes photons to the WGM while maintaining zero pump detuning ($\Delta_c = 0$). The anti-Stokes process encodes phonon information through beam-splitter interactions, enabling indirect detection of molecular vibrations.
For $G/2\pi = 2\,\mathrm{THz}, J_1=J_2=1.5\times 2\pi\,\mathrm{THz}$ and $g_c/2\pi = 50\,\mathrm{GHz}$, this requires optimal pump power $P=8.24\,\mathrm{mW}$ in practical experiment, matching the thermal stable condition of plasmon nanocavity \cite{SsunRevealingPhotothermalBehavior2022}.

\begin{figure}
    \centering
    \includegraphics[width=0.8\textwidth]{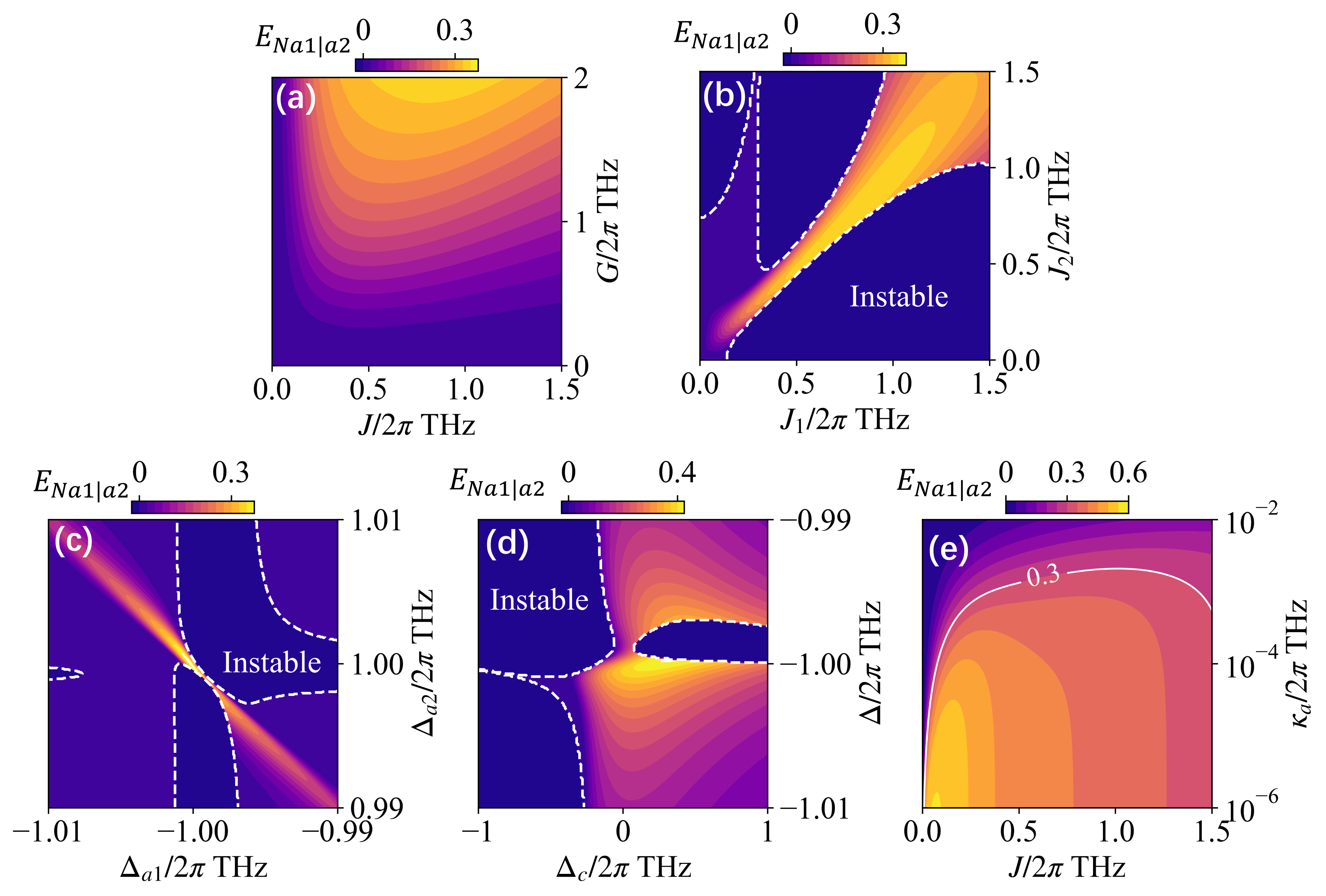}
    \caption{Logarithmic negativity $E_{Na1|a2}$ dependency on different parameters. The initial parameters are $\Delta_{a1}=-\omega_b, \Delta_{a2}=\omega_b, J_1=J_2=2\pi\times1\,\mathrm{THz}, G=2\pi\times 2\,\mathrm{THz},\kappa_a=2\pi \times10^{-3}\,\mathrm{THz}$. Other parameters are the same with Fig.\,\ref{fig:Fig2} in the main text.
    Contour-filled plots of $E_{Na1|a2}$ versus (a) $J_1=J_2=J,G$, (b) $J_1,J_2$, (c) $\Delta_{a1},\Delta_{a2}$, (d) $\Delta_c,\Delta_{a1}=-\Delta_{a2}=\Delta$ and (e) $J_1=J_2=J,\kappa_a$.}
    \label{fig:FigS3}
\end{figure}

The system dynamics are governed by: $$\dot{\hat{u}}(t)=A u(t)+n(t).$$
Note that in our system, the fluctuation operators of two WGMs are both driven by the same zero-point fluctuations in the fiber-guided optical mode, thus the quadrature vectors are: 
\begin{align}
    & \hat u(t)=(\hat X_{a1},\hat Y_{a1},\hat X_b,\hat Y_b,\hat X_c,\hat Y_c,\hat X_{a2},\hat Y_{a2})^T, 
    \\ &\hat n(t)=\sqrt{2}(\sqrt{\kappa_a}\hat X_{a_\mathrm{in}},\sqrt{\kappa_a}\hat Y_{a_\mathrm{in}},\sqrt{\gamma}\hat X_{b_\mathrm{in}},\sqrt{\gamma}\hat Y_{b_\mathrm{in}},\sqrt{\kappa_c}\hat X_{c_\mathrm{in}},\sqrt{\kappa_c}\hat Y_{c_\mathrm{in}},\sqrt{\kappa_a}\hat X_{a_\mathrm{in}},\sqrt{\kappa_a}\hat Y_{a_\mathrm{in}})^T.
\end{align}
And the drifting matrix $A$ is:
\begin{equation}
    A=\left(\begin{matrix}-\kappa_a&\Delta_{a1}&0&0&0&J_1&0&0\\-\Delta_{a1}&-\kappa_a&0&0&-J_1&0&0&0\\0&0&-\gamma&\omega_b&0&0&0&0\\0&0&-\omega_b&-\gamma&G&0&0&0\\0&J_1&0&0&-\kappa_ c&\Delta_c&0&J_2\\-J_1&0&G&0&-\Delta_c&-\kappa_ c&-J_2&0\\0&0&0&0&0&J_2&-\kappa_a&\Delta_{a2}\\0&0&0&0&-J_2&0&-\Delta_{a2}&-\kappa_a\\\end{matrix}\right).
\end{equation}
The covariance matrix $V$ can be solved with $AV+VA^T=-D$, where the diffusion matrix $D$ is:
\begin{equation}
    D=\left(\begin{matrix}\kappa_a&0&0&0&0&0&\kappa_a&0\\0&\kappa_a&0&0&0&0&0&\kappa_a\\0&0&\gamma(2n+1)&0&0&0&0&0\\0&0&0&\gamma(2n+1)&0&0&0&0\\0&0&0&0&\kappa_c&0&0&0\\0&0&0&0&0&\kappa_ c&0&0\\\kappa_a&0&0&0&0&0&\kappa_a&0\\0&\kappa_a&0&0&0&0&0&\kappa_a\\\end{matrix}\right).
\end{equation}
Under symmetric coupling conditions ($J_1 = J_2 = J$) with detunings $\Delta_{a1} = -\omega_b$, $\Delta_{a2} = \omega_b$ and $\Delta_c = 0$, we numerically analyze the photon-photon entanglement $E_{Na1|a2}$. Figure \ref{fig:FigS3}(a) demonstrates maximum entanglement $E_{Na1|a2} \approx 0.34$ at $J/2\pi = 0.87\,\mathrm{THz}$ and $G/2\pi = 2\,\mathrm{THz}$.
Parameter sensitivity analysis reveals:
\begin{enumerate}
    \item \textbf{Coupling asymmetry:} Small discrepancies between $J_1$ and $J_2$ ($<10$\%) induce bearable entanglement degradation (Fig. \ref{fig:FigS3}(b)).
    \item \textbf{Detuning mismatch:} Strict maintenance of $\Delta_{a1} \approx -\Delta_{a2}$ is crucial (Fig. \ref{fig:FigS3}(c)), achievable through thermal compensation of WGMs and pump tuning.
    \item \textbf{Pump detuning tolerance:} Moderate $\Delta_c$ variations preserve entanglement (Fig. \ref{fig:FigS3}(d)).
    \item \textbf{High-Q resonator requirement} While a higher Q factor ($\kappa_a$) enhances the entanglement strength $E_{Na1|a2}$, ultra-high Q values are not strictly necessary. For $\kappa_a/2\pi < 10^{-4}\,\mathrm{THz}$, the improvement in entanglement becomes negligible (Fig. \ref{fig:FigS3}(e)).
\end{enumerate}
Then by measuring the quadrature operators of the output modes of these two WGMs via homodyne detection, the entanglement between the Stokes ($X_s,P_s$) and anti-Stokes ($X_{AS},P_{AS}$) photons in the WGM resonator can be verified via the inseparability criterion: \cite{SduanInseparabilityCriterionContinuous2000}:
    \begin{equation}
        \langle(\Delta X_S+\Delta X_{AS})^2\rangle + 
        \langle(\Delta P_S-\Delta P_{AS})^2\rangle
        <2,
    \end{equation}
which indirectly confirms the presence of WGM-phonon entanglement.

\nocite{*}

\bibliography{supplement}